\documentclass[10pt,letterpaper,twocolumn]{article} 

\usepackage{ol2}
\usepackage[draft]{hyperref}
\usepackage{amsmath}
\usepackage{graphicx}
\usepackage{subfigure}
\usepackage{hyperref}

\begin{document}

\twocolumn[ 

\title{Large nonlinear Kerr effect in graphene}


\author{Han Zhang,$^{1,*}$ St{\'e}phane Virally,$^2$ Qiaoliang Bao,$^3$
Kian Ping Loh,$^3$ \\ Serge Massar,$^{4,*}$  Nicolas Godbout,$^{2}$
and Pascal Kockaert$^1$ }

\address{$^1$OPERA-photonics, Universit{\'e} libre de Bruxelles, 50 Av. F. D. Roosevelt, CP 194/5, B-1050 Bruxelles, Belgium\\
$^2$Engineering Physics Department, \'Ecole polytechnique de Montr\'eal, P.O. Box 6079, Station Centre-ville, \\
Montr\'eal (Qu\'ebec), H3C~3A7 Canada\\
$^3$Department of Chemistry, National University of Singapore, 3 Science Drive 3, Singapore 117543\\
$^4$Laboratoire d'information quantique, CP 225, Universit{\'e} libre de Bruxelles, \\
50 Av. F. D. Roosevelt, B-1050 Bruxelles, Belgium \\
$^*$Corresponding author: hzhang@ulb.ac.be, smassar@ulb.ac.be}

\begin{abstract}
Under strong laser illumination, few-layer graphene exhibits both a
transmittance increase due to saturable absorption and a nonlinear
phase shift. Here, we unambiguously distinguish these two nonlinear
optical effects and identify both real and imaginary parts of the
complex nonlinear refractive index of graphene. We show that
graphene possesses a giant nonlinear refractive index 
\(n_{2}\simeq10^{-7}\,\mathrm{cm^{2}W^{-1}}\), almost nine orders of
magnitude larger than bulk dielectrics. We find that the nonlinear
refractive index decreases with increasing excitation flux but
slower than the absorption. This suggests that graphene may be a
very promising nonlinear medium, paving the way for graphene-based
nonlinear photonics.
\end{abstract}


] 

\noindent Graphene, a single sheet of carbon in a hexagonal lattice,
exhibits many interesting electrical and optical
properties~\cite{propertiesA,propertiesB} which arise due to its
particular bandgap structure $E_{\pm}(\vec{p})=V|\vec{p}|$, where
the sign corresponds to electron (respectively hole) band, $\vec{p}$
is the quasi-momentum, and V$\simeq 10^{6}$~$\mathrm{ms^{-1}}$ is
the Fermi velocity. 
In particular,
a single  graphene layer absorbs a fraction $\pi\alpha\simeq2.3\%$
of the incident light
across the infrared and visible range,
where $\alpha$ is the fine structure constant~\cite{finestructure}.

Graphene also has a broadband ultra-fast saturable
absorption~\cite{Bao2009,SatAbs,Bonaccorso2010,sun2010}, due to
valence band depletion and conductance band filling. This saturable
absorption has been used, amongst other applications, for laser mode
locking~\cite{Bao2009,sun2010}.

\indent The massless bandgap structure of graphene has led to the
prediction of other strong nonlinear optical effects, including
frequency multiplication~\cite{Mikhailov2006, Mikhailov2008} and
frequency mixing~\cite{Mikhailov2011}. Recently, broadband four-wave
mixing in few-layer graphene samples has been
reported~\cite{Hendry2010}. This allowed the determination of the
absolute value of the third order susceptibility of a single
graphene layer $|\chi_{gr}^{(3)}|\simeq 1.5\cdot 10^{-7}$~esu,
approximately eight orders of magnitude larger than in bulk
dielectrics. Four-wave mixing in graphene deposited on optical
ferrules has also been observed~\cite{XuCLEO}.

To our knowledge, there has so far not been any measurement that
unambiguously discriminates between the real and imaginary part of
the third order susceptibility $\chi^{(3)}$ of graphene. We report
herein the experimental measurement of the nonlinear optical
refractive index $n_{2}$ of loosely stacked few-layer graphene using
Z-scan technique. As demonstrated in~\cite{Coso}, the knowledge of
this coefficient together with the nonlinear absorption is
sufficient to calculate the complex value of $\chi^{(3)}$.

\begin{figure}
\centering
\includegraphics[totalheight=5cm]{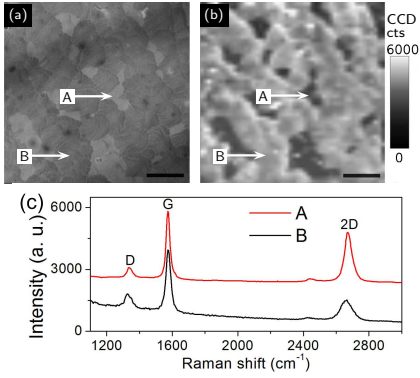}
\caption{Characterization of graphene samples. (a)~Optical image of
graphene sample on silica substrate. Scale bar: 10 $\mu$m. (b)~Raman
map of the G (1560 to 1620~$\mathrm{cm^{-1}}$) band of the area in
(a) (WITec alpha, Laser wavelength: 532~nm,
 spot size: $\sim 500$~nm, 100$\times$ objective). CCD cts.: charge-coupled device counts.
 Scale bar: 10~$\mu$m.
(c)~Raman spectra from the marked spots: A (B) indicates the area
with fainter (darker) contrast given by thinner (thicker) graphene.}
\label{Fig-1}
\end{figure}

Samples of graphene were prepared as follows: a few-layer graphene
thin film was grown on 25~$\mu$m thick Cu foils by chemical vapor
deposition (CVD)~\cite{{LiCaiAn}}. The copper was etched with
ammonium persulfate (0.1~M). After thorough rinsing in distilled
water, the floating graphene film was fished by quartz substrate and
dried gently in nitrogen gas. The optical image of a typical
graphene sample is shown in Fig.~\ref{Fig-1}~(a). The non-uniform
color contrast of the optical micrograph indicates that the film has
a variable thickness, \textit{i.e.}, the dark contrast zones
correspond to thicker graphene domains whereas the faint contrast
zones correspond to thinner graphene domains. This is further
confirmed by the Raman map of the G band, as shown in
Fig.~\ref{Fig-1}~(b), in which bright zones have larger G band
intensity, indicating more graphene layers.
Raman spectra given in
Fig.~\ref{Fig-1}~(c) are captured from marked spots in
Fig.~\ref{Fig-1}~(a), (b), with zone A bilayer and zone B
4-layer graphene. The thickness can be unambiguously
identified by optical contrast (not shown) on optical transmission
measurements.%
Different graphene samples, with respectively 1, 2, 3, 4, and 6 layers
were used in the experiments reported below. The
fabrication process produces loosely stacked graphene layers wherein
the optical properties are expected to arise from the cumulative
contribution of each layer.

We measured the real and imaginary parts of the complex nonlinear
refractive index by using the Z-scan
technique~\cite{Sheikbahae1990}. Our experimental setup has been
used previously in~\cite{Zscan-Moreels}, but has been adapted to the
small thickness of the sample. The sample is subjected to picosecond
pulses emitted from a Pritel picosecond laser with center wavelength
1550~nm, pulse duration 3.8~ps (full width at half maximum), 3-dB
spectral width 1.2~nm and repetition rate 10~MHz. These pulses are
amplified through a Keopsys erbium-doped fiber
amplifier. After out-coupling from the fiber, shaping and routing
the beam through the focusing objective, the average power of the
pulse train can be adjusted between 0 and 4.5~mW. From the laser
parameters, the peak pulse power can therefore be adjusted from 0 to
118~W. The beam is focused using a 20$\times$ microscope objective,
generating a beam waist of 3~$\mu$m, corresponding to illuminations
ranging up to a peak intensity $I$ of $0.84$~$\mathrm{GW\,cm^{-2}}$.
The beam waist was measured by the beam profiler and further
confirmed by z-scan parameter fitting. The graphene sample and its
quartz substrate are oriented perpendicularly to the beam axis and
translated along the axis through the focus with a linear motorized stage. A portion of the amplifier output is picked
off and measured to monitor the optical power level continuously.

\begin{figure}
\centering

\includegraphics[totalheight=5cm]{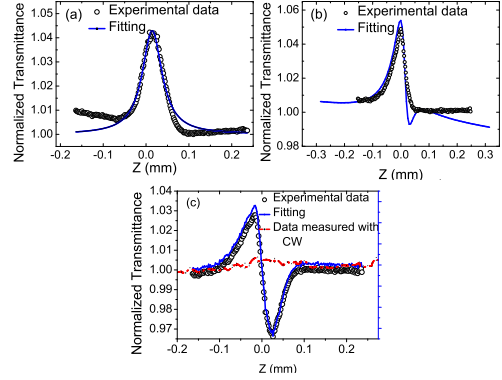}
\caption{Z-scan traces for samples with three-layer graphene taken
at average power of 3~mW, corresponding to a peak power at focus of
0.56~GWcm$^{-2}$. 
(a)~Near field (open aperture). 
(b)~Far field (closed aperture). Upon dividing by the near field curve
one obtains the data of panel (c) which exhibits the typical shape
of a Z-scan curve with positive nonlinear phase shift having an
on-axis value of $\Delta\Phi=0.147$~rad. 
The dotted curve was recorded in cw at the same wavelength and mean power.} 
\label{Fig-2}
\end{figure}

Measurements are performed in two regimes, an open-aperture regime
wherein all the light transmitted through the sample is collected on
a photo-detector and a closed-aperture regime where only an on-axis
portion of the diffracted beam is collected. The open-aperture
regime enables the characterization of the intensity-dependent
absorption. A typical trace when the sample is translated through
the beam focus is shown in Fig.~\ref{Fig-2}~(a). The peak in
transmission when the sample passes through the focus is
characteristic of saturable absorption. The sample's linear
absorption almost completely vanishes for illuminations greater than
0.6~$\mathrm{GW\,cm^{-2}}$. Note that the wings visible in
Fig.~\ref{Fig-2}~(a) present an unusual slope which may arise from
the sample's inhomogeneity. Its presence does not affect our
conclusions.

A typical closed aperture measurement is shown in
Fig.~\ref{Fig-2}~(b). In this trace the effect of the nonlinear
phase is of the same order of magnitude as the effect of saturable
absorption. In order to isolate the former, we divide the curve in
Fig.~\ref{Fig-2}~(b) by the curve in Fig.~\ref{Fig-2}~(a), thereby
obtaining the curve in Fig.~\ref{Fig-2}~(c). The latter has the
typical shape of a Z-scan trace. The up-down curve implies a small
positive on-axis phase shift $\Delta\Phi$. For small $\Delta\Phi$,
one can fit the curve with the function
$T(x)=1+\frac{4x\Delta\Phi}{(1+x^{2})(9+x^{2})}$ where $x=-z/z_{R}$
is the normalized distance from the focus and $z_R$ is the Rayleigh
length. The above method for deriving $\Delta\Phi$ from the measured
data is applicable when the distortion of the phase front is small
and the closed aperture is sufficiently narrow
~\cite{Sheikbahae1990}. We also performed a full calculation of
local nonlinear effects across the graphene layer and its subsequent
propagation, and then fitted Fig.~\ref{Fig-2}~(b) directly. Both
approaches provide comparable results.
Additional measurements
carried out in cw regime with similar average power as in the pulsed
regime ($\simeq 1$~mW) show that our Z-scan  measurements are not
affected by cumulative thermal effects (see Fig.~\ref{Fig-2}(c)). This is presumably due to
the very high thermal diffusion coefficient of graphene.%

\begin{figure}
\centering
\includegraphics[totalheight=5cm]{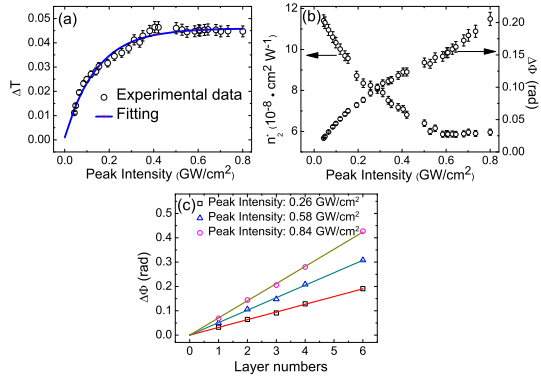}
\caption{ (a)~Relation between transmittance difference ($\Delta T$)
and input power for three layer graphene. (b)~Dependence of
$\Delta\Phi$ (right axis) and $n_2$ (left axis) on peak intensity
for three layer graphene. (c)~Relation between $\Delta\Phi$ and
number of layers, for different peak intensity. } \label{Fig-3}
\end{figure}

Z-scan measurements under variable optical power were performed.
Closed-aperture measurement results show that the transmittance
difference $\Delta T$, which is the difference between the peak and
the background in Fig.~\ref{Fig-2}~(a), has a power dependence
characteristic of saturable absorption. We fitted the transmittance
difference by $\Delta T=\Delta T_{0}-\Delta
T_{0}/(1+I/I_\mathrm{sat})$, see Fig.~\ref{Fig-3}~(a), yielding the
estimate $I_\mathrm{sat}=74$~$\mathrm{MW\,cm^{-2}}$, which is of the
same order of magnitude as the value reported in
~\cite{Martinez2011}, but an order of magnitude smaller than the
values reported in ~\cite{SatAbs}. From this value of
$I_\mathrm{sat}$, we can calculate the nonlinear absorption
coefficient $\beta$ that appears in ~\cite{Coso}. The maximum
transmittance difference for 3-layer graphene is $\Delta
T_{0}=5.1\%$, indicating almost complete saturation of the
absorption. Measurement of the maximal transmittance difference
$\Delta T_{0}$ shows that it scales linearly with the number of
layers (not shown).

The nonlinear phase \(\Delta\Phi\) is plotted on Fig.~\ref{Fig-3}~(b), with respect to the input power. The Kerr refractive index \(n_2\) can be deduced from the slope of this curve at low intensities, using $n_2 = \Delta\Phi/(k_0L I)$ where $k_0=2\pi/\lambda$ and $L$ is the sample thickness,
assumed equal to 1 nm. A value of \(n_{2}\simeq 10^{-7}\,\mathrm{cm^{2}\,W^{-1}}\) is obtained, which is approximately $10^{9}$ times larger than that of bulk dielectrics.
When the intensity is increased, starting from \(I>I_{sat}\approx\mathrm{0.1\,GW\,cm^{-2}}\), the change in the nonlinear phase \(\Delta\Phi\) saturates. In this high intensity regime, the exact modelling of the nonlinear response requires to take into account \(\chi^{(3)}\) and higher order odd terms, such as \(\chi^{(5)}, \ldots\)
However, from the experimental data, an \textit{effective nonlinear index} can be defined by \(n_2^*(I)=\Delta\Phi/(k_0L I)\). One sees that \(n_2^*\) decreases with increasing
$I$, and then reaches a constant value \(n_2^*\approx6\cdot
10^{-8}\,\mathrm{cm^{2}\,W^{-1}}\) for
$I>0.6$~$\mathrm{GW\,cm^{-2}}$.
We also measured the
nonlinear phase as a function of the number of graphene layers and
found that 
at fixed power%
it scales linearly with the number of layers, see
Fig.~\ref{Fig-3}~(c). The real part of the complex
nonlinear refractive index of loosely stacked graphene is found
to be independent of the number of layers.

Note that the
value of $\chi^{(3)}$ based on four-wave mixing experiments reported in~\cite{Hendry2010}
 corresponds to an equivalent $n_2$ of $1.5\cdot
10^{-9}$~$\mathrm{cm^2\,W^{-1}}$, which is forty times smaller than our
value. We attribute this discrepancy to the fact that
 the third order susceptibility of graphene probably has
 multiple origins. The parametric process of four wave mixing measured in~\cite{Hendry2010} arises
 from the coherent electronic response and the massless bandgap structure of graphene, as explained in
Refs.~\citen{Mikhailov2006,Mikhailov2008,Mikhailov2011}. In addition
there is a strong non-parametric process that gives rise to
saturable absorption, and which presumably also gives rise to a
strong non-parametric contribution to $n_2$. For a general discussion of these processes, see~\cite[p.~13]{Boyd2008} 
The Z-scan measurements reported here measure the sum of all contributions to $n_2$.

In conclusion, the characterization of the nonlinear optical
properties of loosely stacked graphene was performed on a Z-scan
setup. Measurements of the saturable absorption are in agreeement
with previously reported values. Measurements of the nonlinear phase
yield a nonlinear coefficient \(n_{2}\simeq 
10^{-7}\,\mathrm{cm^{2}\,W^{-1}}\), where $n_2$ is defined
in~\cite{Coso}.  
The measured value of the effective nonlinear refractive index \(n_2^*\) decreases with
increasing input power. Interestingly it remains large when the
absorption is saturated. This suggests that at high power
(\(I>0.6\,\mathrm{GW\,cm^{-2}}\)) the figure of merit of graphene for
nonlinear optics applications may be highly favorable (high
nonlinear phase, low absorption per layer).

This work is funded by the Belgian
Science Policy Office (BELSPO) Interuniversity Attraction Pole (IAP)
programme under grant no.~IAP-6/10 (HZ, SM, PK), by the LKY
Postdoctoral Fellowship (QLB), by the NRF-CRP Graphene Related
Materials and Devices (KPL), and by the National Science and
Engineering Research Council of Canada (SV, NG).

\pagebreak

\section*{Informational Fourth Page}
In this section, please provide full versions of citations to
assist reviewers and editors (OL publishes a short form of
citations) or any other information that would aid the peer-review
process.

\end{document}